\documentclass[pra,aps,showpacs,superscriptaddress,twocolumn]{revtex4}
\usepackage{graphicx}
\usepackage{amsmath}
\usepackage{amsfonts}
\usepackage{amssymb}
\usepackage{epsfig}

\begin{document}
\title{
Entanglement of a qubit coupled to a resonator in the adiabatic
regime}
\author{Giuseppe Liberti}\email{liberti@fis.unical.it}\affiliation{
Dipartimento di Fisica, Universit\`a della Calabria, 87036
Arcavacata di Rende (CS) Italy}
\author{Rosa Letizia
Zaffino}\affiliation{ Dipartimento di Fisica, Universit\`a della
Calabria, 87036 Arcavacata di Rende (CS) Italy} \affiliation{INFN
- Gruppo collegato di Cosenza, 87036 Arcavacata di Rende (CS)
Italy}
\author{Francesco Piperno}\affiliation{
Dipartimento di Fisica, Universit\`a della Calabria, 87036
Arcavacata di Rende (CS) Italy}
\author{Francesco Plastina}\affiliation{ Dipartimento di Fisica, Universit\`a della
Calabria, 87036 Arcavacata di Rende (CS) Italy} \affiliation{INFN
- Gruppo collegato di Cosenza, 87036 Arcavacata di Rende (CS)
Italy}
\date{\today}

\begin{abstract}
We discuss the ground state entanglement of a bi-partite system,
composed by a qubit strongly interacting with an oscillator mode,
as a function of the coupling strenght, the transition frequency
and the level asymmetry of the qubit. This is done in the
adiabatic regime in which the time evolution of the qubit is much
faster than the oscillator one. Within the adiabatic
approximation, we obtain a complete characterization of the ground
state properties of the system and of its entanglement content.
\end{abstract}
\bigskip
\pacs{03.67.Mn,03.65.Ud,03.65.Yz} \maketitle
\section{Introduction and description of the model}
The spin-boson model has been widely used to investigate the
interaction between a two-level system, a qubit, and an harmonic
oscillator environment, describing fluctuations of either
electromagnetic or elastic origin \cite{weiss}. The coupling of
the qubit with each environmental mode gives rise to a progressive
entanglement, leading to the decoherence of the qubit itself.

This model has been largely employed in the weak coupling limit to
explain noise effects  in solid state devices which could be
useful for quantum information processing \cite{yuma}. It has been
also applied to describe the coupling of such devices to quantum
detectors.

In the latter perspective, the strong coupling to a single bosonic
mode has been analyzed in Ref. \cite{levine}. This kind of
``restriction'' to a single-mode environment appears to be useful
for the decoherence problem too, as recent experimental and
theoretical works have attributed a prominent role to the coupling
of superconducting Josephson qubits with spurious micro-resonator
resulting from the presence of switching charged impurities
residing in the tunnel barrier\cite{martinis,pino}.

In this paper, we analyze the case of a qubit strongly coupled to
a slow resonator, working in the adiabatic regime. Our model is a
generalization of the one employed in Ref. \cite{levine}, which
turned out to also describe molecular Jahn-Teller effect
\cite{hines}. It can be also used to describe the coupling of a
Josephson charge qubit to an electromagnetic resonator
\cite{blais} or to another (large) junction working in the
harmonic regime\cite{prb03}, in the case of strong and
off-resonant interaction. As it occurs for many solid state
implementations, we assume, here, that the coupling can become so
strong that the usual rotating wave approximation cannot be
employed.

Our aim is to characterize the ground state of the system and, in
particular, to evaluate the amount of quantum correlation present
in (that is, the ``entanglement content'' of) the fundamental
level. If the presence of the oscillator is spurious and
un-wanted, this ``residual'' entanglement can produce errors in
the information processing performed by the qubit. An
investigation of the entanglement in the case of a two state
system coupled to an ohmic environment has been performed by Costi
and McKenzie \cite{costi}, by exploiting the equivalence to the
anisotropic Kondo model. They were able to show that the
entanglement entropy, for level asymmetry different from zero,
reaches a maximum at smaller values of dimensionless dissipation
strength.

In fact, the calculation of ground state entanglement has been
used to characterize complex quantum many body systems, with
particular emphasis on its connection to quantum phase transitions
\cite{ent1,ent2,ent3,ent4}. In our case, the system is bi-partite,
and therefore there is no collective behavior to be examined; but
nevertheless, a kind of criticality is present, as in the massive
limit for the oscillator (and for qubit working at degeneracy),
two regions in parameter space exist, exhibiting completely
separable and entangled ground state, respectively \cite{levine}.
Furthermore, a sharp increase from zero is found at the onset of
entanglement. This has been interpreted as a quantum reminiscence
of the bifurcation of the fixed point of the oscillator in the
corresponding classical model \cite{hines}.

We show below that, within the adiabatic approximation scheme,
this behavior can be obtained analytically together with the
leading corrections for a finite tunnelling amplitude of the
qubit. Our approach, however, is not limited to this region and we
show that it can be used to systematically investigate ground
state properties and entanglement in a broader parameter range, as
we can account also for the effect of level asymmetry.

\vskip 12pt A qubit interacting with a single harmonic oscillator
mode can be
 described by the Hamiltonian (in unit such that $\hbar=c=1$)
\begin{equation}
    H={\Delta}\sigma_x+\left[{\epsilon} +
   \frac{\lambda}{\sqrt{2m\omega}}(a^{\dagger}+a)\right]\sigma_z+\omega a^{\dagger}a
    \label{1}
    \end{equation}
where $\Delta$ is the transition frequency of the qubit,
$\epsilon$ is the level asymmetry, $\omega$ is the frequency of
the oscillator and $\lambda$ is the coupling strength.

Hines et al. \cite{hines} start their description from the case of
frozen qubit (i.e., $\Delta=\epsilon=0$), which allows for an
exact solution of the problem. Indeed, the Hamiltonian has doubly
degenerate eigenstates which can be represented by displaced
oscillator states \cite{crisp}. In this degeneracy limit, one
obtains two displaced harmonic oscillator wells, with equilibrium
positions $q_0=\pm \lambda/m\omega^2$, so that the qubit can be
localized in either the left or the right well. The system's wave
function can be expanded in terms of a complete set of these
orthonormal states and, for all superposition of the two
degenerate levels, the ground state entanglement is found to
increase with the coupling, becoming maximal as
$\lambda\rightarrow\infty$. For $\Delta\neq0$ (but $\epsilon=0$),
a numerical analysis is then performed in the same basis, with
some necessary truncation of the Hilbert space of the system. This
approximation is known as ``quasi-degenerate limit'' and it has
been shown to provide an accurate description of the system only
for $\Delta/\omega\leq1$, see Ref. \cite{Irish}, due to the lack
of orthogonality between different displacements.

\vskip 12pt Here, we work in the opposite regime, and assume a
{\it fast} qubit, $\Delta \gg \omega$, to perform the adiabatic
approximation as described in the following section. Ground state
entanglement is evaluated in section \ref{sec3}, while section
\ref{sect4} gives some concluding remarks. \section{Adiabatic
approach}\label{sect2}The Born-Oppenheimer approximation scheme
can be followed more plainly by rewriting the Hamiltonian of Eq.
(\ref{1}) as follows
\begin{equation}
   H=\frac{\omega}{2}\left[Q^2+P^2+D \sigma_x+\left(W
   +L Q\right)\sigma_z\right] \, ,
    \label{hc}
\end{equation}
where the oscillator coordinates representation has been
introduced,
\begin{equation}\label{coordre}
    Q = \frac{1}{\sqrt{2}}(a^{\dagger}+a)\, , \qquad
    P=i{\frac{1}{\sqrt{2}}}(a^{\dagger}-a) \, ,
\end{equation}
together with the dimensionless parameters $D=2\Delta/\omega$,
$W=2\epsilon/\omega$ and $L=2 \lambda / \sqrt{m \omega^3}$.

The basic assumption of the well-known adiabatic approximation is
that the total wave function  of a composite system with one fast
(the qubit) and one slowly (the oscillator) changing part can be
written as:
\begin{equation}\label{deco}
    |\psi_{tot}\rangle=\int d Q \, \phi(Q) |Q\rangle \otimes |\chi (Q)\rangle
\end{equation}
The states $|\chi (Q)\rangle$ are the eigenstates of the
``adiabatic'' equation of the qubit part for each fixed value of
the slow variable $Q$,
\begin{equation}\label{adiaham}
\left[D \sigma_x+\left(W +L Q \right ) \sigma_z \right]
|\chi_\sigma(Q)\rangle=E_\pm(Q)|\chi_\sigma(Q)\rangle \,,
\end{equation}
which gives the eignvalues
\begin{equation}\label{dq}
E_\pm(Q)=\pm E(Q)=\pm\sqrt{D^2+(W+LQ)^2}\,.
\end{equation}
The two eigenstates of Eq. (\ref{adiaham}) can be written as
\begin{eqnarray}\label{gsdgen}
|\chi_l(Q)\rangle&=&\frac{1}{\sqrt{2}}\left(A_-(Q)|+\rangle-A_+(Q)|-\rangle\right) \, ,\\
|\chi_u(Q)\rangle&=&\frac{1}{\sqrt{2}}\left(A_+(Q)|+\rangle+A_-(Q)|-\rangle\right)
\, ,
\end{eqnarray}
where $|\pm\rangle$ are the eigenstates of $\sigma_z$ with
eigenvalues $\pm1$ and
\begin{equation}\label{gsdgen2}
    A_{\pm}(Q)=\sqrt{{1\pm\frac{W+LQ}{E(Q)}}} \, .
\end{equation}
The subscripts $l$ and $u$ refers to the lower and to the upper
effective adiabatic potentials felt by the slow oscillator,
respectively,
\begin{equation}\label{udq}
    U_{u,l}(Q)=\frac{\omega}{2}\left[Q^2\pm E(Q)\right] \, .
\end{equation}
As we are primarily interested on ground state properties, we will
concentrate on $U_l$ from now on.

A special case of interest is the one with the qubit working at
degeneracy, $W=0$. In this case one obtains a symmetric
Hamiltonian with conservation of total parity (given by the joint
transformation $Q\rightarrow-Q$ and
$\sigma_z\rightarrow-\sigma_z$). Introducing the dimensionless
parameter
\begin{equation}\label{dp}
\alpha=\frac{L^2}{2 D}=\frac{\lambda^2}{m\omega^2\Delta}\,,
\end{equation}
one can show that for $\alpha\leq1$, the potential
$U_{l}^{W=0}(Q)$ can be viewed as a broadened harmonic potential
well with its minimum at $Q=0$ and $U_{l}^{W=0}(0)=-\Delta$. For
$\alpha>1$, on the other hand, the coupling of the oscillator with
the qubit splits the oscillator potential producing a symmetric
double well with the minima at
\begin{equation}\label{ground}
    Q=\pm Q_0=\pm\frac{D}{L}\sqrt{\alpha^2-1} \, ,
\end{equation}
with
\begin{equation}\label{eq0}
 U_{l}^{W=0}(\pm
 Q_0)=-\frac{\Delta}{2}\left(\alpha+\frac{1}{\alpha}\right)\, .
\end{equation}
$Q_0$ is used as a kind of order parameter in Ref. \cite{levine},
in the limit $D \rightarrow \infty$.

\begin{figure}
\includegraphics{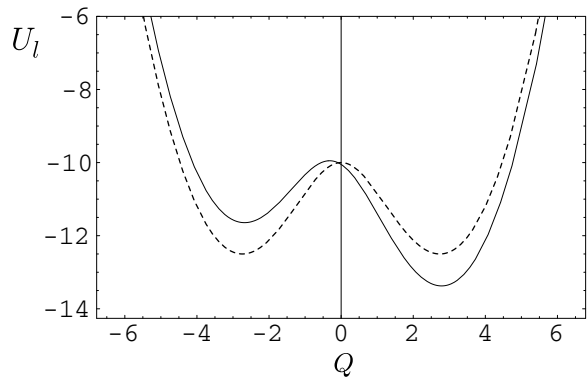}\\
\caption{\label{pot} The lower adiabatic potential for $D=10$ and
$\alpha=2$. The dashed line refers to the symmetric, $W=0$, case
(dashed line), while the solid line refers to $W=1$. The case of
frozen qubit ($W=D=0$) would have given a pair of independent
parabolas instead of the adiabatic potentials $U_{l,u}$ of Eq.
(\ref{udq}).}
\end{figure}

For $W\neq0$, the symmetry is broken, and for this reason we refer
to $W$ as the asymmetry parameter. The form of lower potential for
two different sets of parameters is shown in Fig.(\ref{pot}).

Having obtained the state of the qubit, the last step in the
adiabatic procedure is now to evaluate the ground state wave
function for the oscillator, $\phi_0(Q)$, to be inserted in Eq.
(\ref{deco}) to obtain the fundamental level of the coupled
system. This wave function satisfies the one-dimensional time
independent Schrodinger equation
\begin{equation}
H_{ad} \, \phi_{0}(Q)=\left(-\frac{\omega}{2}\frac{d^2}{dQ^2}+
U_{l}(Q)\right)\phi_{0}(Q)=E_{0}\phi_{0}(Q) \label{hc22} \, ,
\end{equation}
with $\int_{-\infty}^\infty \phi_{0}^2(Q) dQ=1 $, and where $E_0$
is the lowest eigenvalues of the adiabatic Hamiltonian defined by
the first equality.

\begin{figure}
\includegraphics{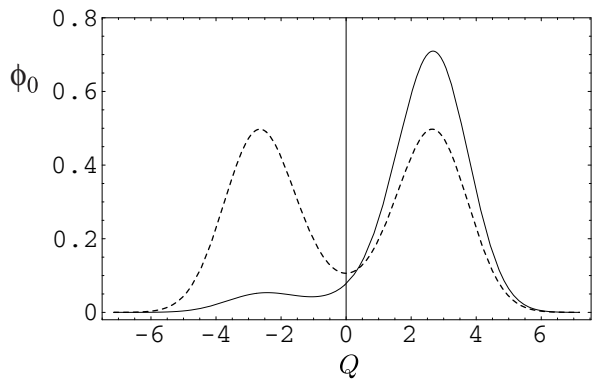}\\
\caption{\label{wf} Normalized ground state wave function for the
oscillator in the lower adiabatic potential, for $D=10$ and
$\alpha=2$ and with $W=0$ (dashed line) and $W=0.1$ (solid line).}
\end{figure}

In Fig.(\ref{wf}), the wave function $\phi_0(Q)$ is shown with
$D=10$ and $\alpha=2$ for both the degenerate, $W=0$, and a
slightly asymmetric case, $W=0.1$.
Notice that even a very small value of the asymmetry parameter $W$
gives rise to a wave function almost localized in the lower well.

Given the function $\phi_0(Q)$, we can evaluate the reduced
density matrix for the qubit and obtain the ground state
entanglement. This is done in the next section.

\section{Reduced qubit state and entanglement}\label{sec3}
\begin{figure}
 \includegraphics{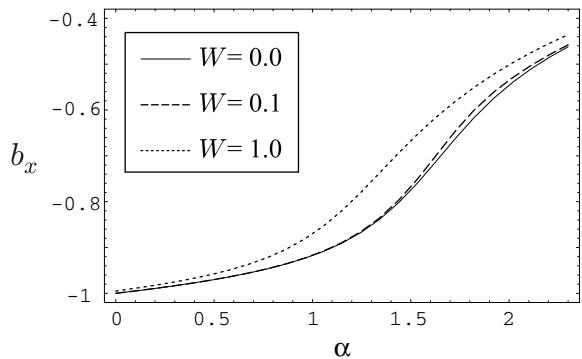}\\
 \caption{\label{asx} The dependence of the ground state expectation value
$b_x=\langle\sigma_x \rangle$ as a function of the parameter
$\alpha$, for $D=10$.}
\end{figure}

\begin{figure}
\includegraphics{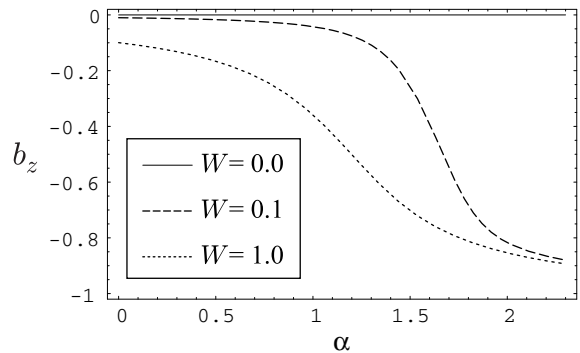}\\
\caption{\label{asz} The $z$-component of the Bloch vector as a
function of $\alpha$, for $D=10$.}
\end{figure}

The reduced density operator describing the qubit alone, when the
overall  system is in the ground state can be written as
\begin{equation}\label{rmdpm}
\rho_{0}=\int_{-\infty}^\infty dQ \,
|\psi_0(Q)\rangle\langle\psi_0(Q)|=\frac{1}{2}\left(I+ b_x
\sigma_x+ b_z \sigma_z\right) \,
\end{equation}
where $\vec b = \langle \vec{\sigma} \rangle$ is the Bloch vector,
whose non-zero components are explicitly given by the following
integrals
\begin{equation}\label{sxm}
b_x=-\int_{-\infty}^\infty \phi_{0}^2(Q) \frac{D}{E(Q)} dQ \, ,
\end{equation}
and
\begin{equation}\label{szm}
b_z= -\int_{-\infty}^\infty \phi_{0}^2(Q) \frac{W+LQ}{E(Q)} dQ \,
. \end{equation} In Fig.(\ref{asx}) and (\ref{asz}), we show the
dependence on the dimensionless quantity $\alpha=L^2/2 D$ of the
ground state expectation values defined by Eqs. (\ref{sxm}) and
(\ref{szm}), respectively.

It is easily seen that $b_z$ is different from zero only for $W\ne
0$, while in the symmetric case the population is equally
distributed between the states $|\pm\rangle$ of the qubit. This is
due to the inversion symmetry of the adiabatic potential, which,
for finite $D$, implies that the system does not localize in any
of the wells and, consequently, that the state of qubit does not
have a well defined value of $\sigma_z$.

In the case $W=0$, the integrand of Eq. (\ref{sxm}) becomes the
product of the squared ground state wave function and the square
root of a Lorentz function centered at $Q=0$. In the limit
$\alpha\rightarrow0$, this integral reduces to the normalization
condition for the ground state wave function and thus
$b_x\simeq-1$. In fact, it is possible to show that, for small
$\alpha$, the main effect of the qubit is to renormalize the value
of the oscillator frequency by a factor $k=\sqrt{1-\alpha}$. As a
result, the adiabatic ground state wave function for the
oscillator is approximately given by
\begin{equation}
\phi_0(Q) \simeq \left ( \frac{k}{\pi} \right )^{\frac{1}{4}} \,
\exp \left \{ - \frac{k}{2} \, Q^2 \right \} \, ,
\end{equation}
so that
\begin{equation} b_x \simeq -1 + \frac{\alpha}{2 D k} \, , \quad
\mbox{for } \, \alpha \ll 1 \; .\end{equation} For $\alpha\gg1$
the ground state wave function is located in spatial regions far
from $Q=0$ and thus $b_x\simeq0$. To obtain an analytic estimation
for large $\alpha$, we can take as an approximate adiabatic ground
state for the oscillator the symmetric superposition
\begin{equation}
\phi_0(Q) \simeq \frac{1}{\sqrt{2}} \, \left  \{ \phi_+(Q) +
\phi_-(Q) \right \} \, ,
\end{equation} with
\begin{equation}
\phi_{\pm} (Q) = \left ( \frac{k'}{\pi} \right )^{\frac{1}{4}} \,
\exp \left \{- \frac{k'}{2} (Q\mp Q_0)^2 \right \} \, ,
\label{due1}\end{equation} where $k'= 1- \frac{1}{D \alpha^2}$ is,
again, a renormalization factor for the oscillator frequency.

\noindent Taking the dominant contribution in Eq. (\ref{sxm}), one
gets
\begin{equation}
b_x \simeq - \frac{1}{\alpha} - \frac{2}{D \alpha^2} \qquad
\mbox{for } \, \alpha \gg 1 \, , \label{expa}\end{equation} which
we checked to be in very good agreement with the numerical
solution.

For $W\neq 0$, $b_z$ becomes non-zero and decreases monotonically
with increasing $\alpha$ with $-1$ as limiting value for
$\alpha\gg1$. This is due to the fact that the $\sigma_z$
contribution dominates in the qubit Hamiltonian in this regime,
and therefore the qubit stays in the state $|-\rangle$.

Even in the presence of an asymmetry $W$, the $x$ component of the
Bloch vector continues to grow monotonically from $-1$ to $0$ when
$\alpha$ increases from zero, with only quantitative deviation
from the $W=0$ behavior.

We can, thus, summarize by saying that for small $\alpha$, the
state of the qubit is the lower eigenstate of $\sigma_x$, while
for large enough $\alpha$ the qubit is found to be in the lower
eigenstate of $\sigma_z$. Between these two extreme cases, a
cross-over occurs, which becomes a true, sharp transition for very
large $D$ and $W=0$ (see below).

\begin{figure}
\includegraphics{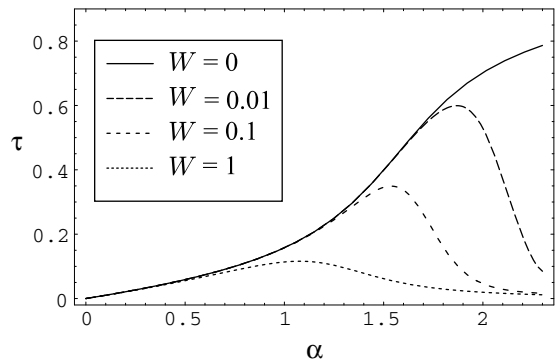}\\
\caption{\label{tau10} The tangle $\tau$ as a function of $\alpha$
for $D=10$. Different curves, corresponding to different values of
$W$, indicate the entanglement progressively decreases with
increasing the asymmetry.}
\end{figure}

\begin{figure}
\includegraphics{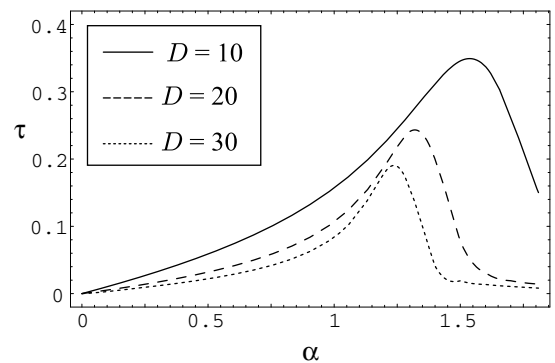}\\
\caption{\label{tau01} The tangle $\tau$ as a function of $\alpha$
for $W=0.1$ and different values of $D$.}
\end{figure}

\begin{figure}
\includegraphics{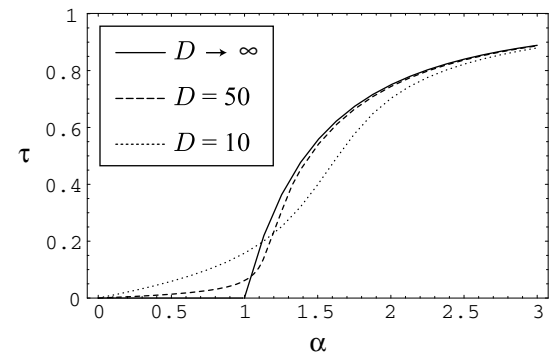}\\
\caption{\label{tau0} The tangle as a function of $\alpha$ in the
symmetric case $W=0$ for different values of the qubit tunnelling
amplitude $D$. One can appreciate that the result of Eq.
(\ref{tangl}) is indeed reached asymptotically. }
\end{figure}

The knowledge of the qubit reduced density matrix allows us to
evaluate the entanglement in the ground state. A quantitative
measure of the entanglement between the qubit and the oscillator
is given by the \emph{tangle} \cite{Rungta}, which, for globally
pure states, is defined as
\begin{equation}\label{tau}
    \tau=2\left[1-{\mathrm{Tr}}(\rho_0^2)\right] \, .
\end{equation}
This is an entanglement monotone, giving $\tau=0$ for a separable
state and reaching $\tau =1$ for maximally-entangled states. In
our case, Eq. (\ref{tau}) becomes
\begin{equation}\label{tau2}
    \tau=1-b_x^2-b_z^2
\end{equation}
The tangle  is shown in Figs. (\ref{tau10}), (\ref{tau01}) and
(\ref{tau0}) for different values of the $D$ and $W$ parameters,
as a function of the dimensionless quantity $\alpha$. For any
$W\neq0$, the entanglement increases with increasing $\alpha$
before reaching a maximum value; after that, it decreases again to
zero for $\alpha\gg1$.  As stated above, this is due to the fact
that the state of the system factorizes in this limit if $W \neq
0$. In the symmetric case, see Fig.(\ref{tau0}), the entanglement
becomes maximal as the coupling increases and, in the strict
adiabatic limit $D\rightarrow\infty$, the tangle becomes
discontinuous at the critical value $\alpha=1$ and rapidly
increases from zero to one when $\alpha> 1$.

This result has a simple analytic derivation that is easily
obtained from the thermal ground-state of the system. The reduced
density operator describing the ground state of the qubit may be
found by tracing out the oscillator variables from the thermal
state
\begin{equation}\label{zts}
    \rho=\frac{e^{-\beta H}}{Z(\beta)} \, ,
\end{equation}
and by taking the limit $\beta\rightarrow\infty$. Here
$Z(\beta)={\mathrm{Tr}}\left\{e^{-\beta H}\right\}$ is the
partition function.  The thermal density $\rho$ possesses the full
symmetry as the Hamiltonian $H$ and, if the ground state is non
degenerate, the zero-temperature state coincides with the ground
state of the system. It is important to stress that this is not
generally true for degenerate ground states (as in the case
$\Delta=\epsilon=0$). When a degeneracy arises, each individual
ground state may not possess the same symmetries of the
Hamiltonian. Instead, they are always shared by the zero
temperature state, which is just an equal mixture of all the
possible ground states. This situation does not occur in our case.

\vskip 12pt By rewriting the Hamiltonian (\ref{hc}) as
\begin{equation}
   H=\frac{1}{2 m}p^2+\frac{m\omega^2}{2}q^2+{\Delta} \sigma_x
   +\left({\epsilon}+\lambda q\right)\sigma_z \, ,
    \label{hcmassive}
    \end{equation}
where
\begin{equation}\label{coordremass}
    q=\frac{1}{\sqrt{m\omega}}Q\,,\quad    p=\sqrt{{m\omega}}P \,
    ,
\end{equation}
we see that the limit $m\rightarrow\infty$ and
$m\omega^2\rightarrow$ const. is equivalent to neglect the kinetic
energy of the oscillator.  In this regime, thus, one gets
\begin{eqnarray}
Z(\beta)& =& {\mathrm{Tr}}\int_{-\infty}^\infty dq \langle
q|e^{-\frac{\beta m\omega^2 }{2}q^2}e^{-\beta\left({\Delta}
\sigma_x+ \left({\epsilon}+\lambda q\right)\sigma_z\right)}|q\rangle\nonumber\\
& =& 2\int_{-\infty}^\infty dq e^{-\frac{\beta m\omega^2
}{2}q^2}\cosh{\left[\beta\sqrt{{\Delta^2}+
{\left({\epsilon}+\lambda q\right)^2}}\right]}\nonumber\\
\label{Z}
\end{eqnarray}
In the first row, the states $|q\rangle$, employed to perform the
trace, are just the position eigenstate of the oscillator.

The thermal reduced density matrix of the qubit can be formally
written in the form of Eq.(\ref{rmdpm}) and is obtained by tracing
out the thermal reduced density matrix over the oscillator degree
of freedom. The temperature-dependent expectation values of
$\sigma_x$ and $\sigma_x$, respectively, are, then
\begin{equation}
b_x= -\frac{2}{Z(\beta)}\int_{-\infty}^\infty dq
\frac{\Delta}{\Delta(q)} e^{-\frac{\beta m\omega^2
    }{2}q^2}\sinh{\left[\beta\Delta(q)\right]} \, ,
\end{equation}
\begin{equation}
b_z= -\frac{2}{Z(\beta)}\int_{-\infty}^\infty dq
\frac{{\epsilon}+\lambda q}{\Delta(q)}
    e^{-\frac{\beta m\omega^2
    }{2}q^2}\sinh{\left[\beta\Delta(q)\right]}\, ,
\end{equation}
where $\Delta(q)=\sqrt{\Delta^2+({\epsilon}+\lambda q)^2}$.

We focus, again, our discussion on the ground state
($\beta\rightarrow\infty$). In this limit, the partition function
may be evaluated by the steepest descent method. For $\epsilon=0$,
the integrand of Eq.(\ref{Z}) is symmetric around $q=0$. When
$\beta\rightarrow\infty$ and $\alpha\leq1$ this function has only
a sharp maximum at the origin, while for $\alpha>1$, the integrand
has two sharp maxima at $q=\pm \Delta\sqrt{\alpha^2-1}/\lambda$,
symmetrically displaced around zero, where the function has a
shallow minimum. In this limit, one easily obtain $b_z=0$ and
\begin{equation}
b_x=\left\{%
\begin{array}{ll}
    -1, & \hbox{$\alpha\leq 1;$} \\
    \hbox{$-1/\alpha$}, & \hbox{$\alpha> 1.$} \\
\end{array}%
\right.
\end{equation}
Then, for the tangle, one simply gets
\begin{equation}
    \tau=\left\{%
\begin{array}{ll}
    0, & \hbox{$\alpha\leq 1;$} \\
    \hbox{$1-1/\alpha^2$}, & \hbox{$\alpha> 1.$} \\
\end{array}%
\right. \label{tangl} \, .
\end{equation}
The first $1/D$-correction to this result can be obtained, for
large $\alpha$ by using the expansion given in Eq. (\ref{expa})
together with the definition of the tangle, Eq. (\ref{tau2}).
These results are shown in Fig. (\ref{tau0}), where the solid line
is a plot of the tangle in the asymptotic regime ($D \rightarrow
\infty$).

The same procedure used above can be carried out in the asymmetric
case provided a value of $q$ is found, such that
\begin{equation}\label{smq}
{q}=\frac{\lambda}{m\omega^2}\frac{\epsilon+\lambda
{q}}{\sqrt{\Delta^2+(\epsilon+\lambda {q})^2}} \, .
\end{equation}
This equation has three nontrivial solutions. Within our
saddle-point scheme, in the limit $\beta\rightarrow\infty$, we
must retain only the solution ${q}_m$ that corresponds to the
lowest minimum of the potential. Therefore, we can write
\begin{equation}\label{ahia}
b_x= - \frac{\Delta}{\Delta(q_m)}\,,\quad b_z= -
\frac{\epsilon+\lambda q_m}{\Delta(q_m)} \, ,
\end{equation}
and, thus, one gets $\tau=0$ for any finite $\epsilon\neq0$.

Indeed, it can be seen from Fig. (\ref{tau10}) that the tangle
(for any value of $\alpha$) decreases progressively with the
increase of the asymmetry parameter $W$. Furthermore, as
exemplified in Fig. (\ref{tau01}), for any fixed non-zero value of
$W$, the tangle approaches zero as $D$ increases; so that,
asymptotically, the result implied by Eq. (\ref{ahia}) is
obtained.
\section{Concluding remarks}\label{sect4}
In conclusion, we have discussed the adiabatic approximation for a
qubit coupled to a single oscillator mode and we have derived the
resulting entanglement in the ground state, by giving simple
analytical results in the strict adiabatic limit. The advantage of
our approach, that requires a very small computational effort and
correctly describes the model system when $\Delta/\omega\gg1$, is
to give a physically more transparent description of the ground
state.

As we have shown, the procedure is easily extended to the
asymmetric case and this is important since the entanglement
changes dramatically for any finite (however small) value of the
asymmetry in the qubit Hamiltonian. As mentioned in section
\ref{sect2} above, this is due to the fact the this term modifies
the symmetry properties of the Hamiltonian, so that the form of
the ground state changes radically and the same occurs to the
reduced qubit state. For example, for a large enough interaction
strength, the qubit state is a complete mixture if $W=0$, while it
becomes the lower eigenstate of $\sigma_z$ if $W \ne 0$. As a
result, for large $\alpha$, there is much entanglement if $W=0$,
while the state of the system is factorized and thus $\tau=0$ if
$W \ne 0$. This is seen explicitly in Fig. (\ref{tau10}).
Furthermore, from the comparison of Figs. (\ref{tau10}),
(\ref{tau01}), and ({\ref{tau0}), one can see that, with
increasing $\alpha$, the tangle increases monotonically in the
symmetric case, while it reaches a maximum before going down to
zero if $W\ne 0$.

This is due to the fact that, in the first case, the ground state
of the system becomes a Schr\"{o}dinger cat-like entangled
superposition, approximately given by
\begin{equation}
|\psi \rangle \approx \frac{1}{\sqrt 2} \Bigl \{ |\phi_+\rangle
|-\rangle - |\phi_- \rangle |+ \rangle \Bigr \} \, , \quad
\mbox{for } \alpha \gg 1 \, , \label{schroca} \end{equation} where
$|\phi_{\pm} \rangle$ are the two coherent states for the
oscillator defined in Eq. (\ref{due1}), centered in $Q= \pm Q_0$,
respectively, and almost orthogonal if $\alpha \gg 1$.

In the presence of asymmetry, on the other hand, the oscillator
localizes in one of the wells of its effective potential and this
implies that, for large $\alpha$, the ground state is given by
just one of the two components superposed in Eq. (\ref{schroca}).
This is, clearly, a factorized state and therefore one gets
$\tau=0$.

Since $\tau$ is zero for uncoupled sub-systems (i.e., for very
small values of $\alpha$), weather $W=0$ or not, and since, for
$W\ne 0$, it has to decay to zero for large $\alpha$ , it follows
that a maximum is present in between.

In fact, for intermediate values of the coupling,  there is a
competition between the $\alpha$-dependences of the two non zero
components of the Bloch vector. In particular, the length $|\vec
b|$ is approximately equal to one for both small and large
$\alpha$'s, see Figs. (\ref{asx})-(\ref{asz}), but the vector
points in the $x$ direction for $\alpha \ll 1$ and in the $z$
direction for $\alpha \gg 1$. The maximum of the tangle in the
asymmetric case occurs near the point in which $b_x \approx b_z$.

For the symmetric case, we were also able to derive analytically
the sharp increase of the entanglement at $\alpha=1$. This
behavior appears to be reminiscent of the super-radiant transition
in the many qubit Dicke model, which, in the adiabatic limit,
shows exactly the same features described here, and which can be
described along similar lines.

Finally, we would like to comment on the relationship of this work
with those of Refs. \cite{levine} and \cite{costi}. The approach
proposed by Levine and Muthukumar, Ref. \cite{levine}, employs an
instanton description for the effective action. This has been
applied to obtain the entropy of entanglement in the symmetric
case, in the same critical limit described above. It turns out
that this description is equivalent to a fourth order expansion of
the lower adiabatic potential $U_l$. This approximation, although
retaining all the distinctive qualitative features discussed
above, gives slight quantitative changes in the results.

Concerning the asymmetric case, our results for the ground state
entanglement appear similar to those found by Costi and McKenzie
in Ref. \cite{costi}, where the interaction of a qubit with an
ohmic environment was numerically analyzed. It turns out that, for
a bath with finite band-width, the entanglement displays a
behavior analogous to that reported in Figs.
(\ref{tau10})-(\ref{tau01}), when plotted with respect to the
value of the impedance of the bath. Here, instead, we concentrated
on the dependence of the tangle on the coupling strength between
the qubit and the environmental oscillator. Unfortunately, the
coupling strength is not easily related to the coefficient of the
spectral density used in Ref. \cite{costi}, and therefore one
cannot make a precise comparison between the two results. At least
qualitatively, however, we can say that the ground state quantum
correlations induced by the coupling with an ohmic environment are
already present when the qubit is coupled to a single oscillator
mode.

\end{document}